\documentstyle[prd,aps,twocolumn,psfig]{revtex}

 \newcommand{\CL}{{\cal L}}

\newcommand{\bear}{\begin{array}}  \newcommand{\eear}{\end{array}}
\newcommand{\bea}{\begin{eqnarray}}  \newcommand{\eea}{\end{eqnarray}}
\newcommand{\beq}{\begin{equation}}  \newcommand{\eeq}{\end{equation}}
\newcommand{\bef}{\begin{figure}}  \newcommand{\eef}{\end{figure}}
\newcommand{\bec}{\begin{center}}  \newcommand{\eec}{\end{center}}
\newcommand{\non}{\nonumber}  
\newcommand{\lmk}{\left(}  \newcommand{\rmk}{\right)}

\newcommand{\del}{\partial}  
\newcommand{\vect}[1]{\mbox{\boldmath${#1}$}}
\newcommand{\bib}{\bibitem} \newcommand{\new}{\newblock}
\newcommand{\la}{\left\langle} \newcommand{\ra}{\right\rangle}

\newcommand{\vex}{\mbox{\boldmath${x}$}}

\def\IB#1#2#3{{\bf #1}, #2 (19#3)}
\def\IBB#1#2#3{{\bf #1}, #2 (20#3)}
\def\IBID#1#2#3{{\it ibid}. {\bf #1}, #2 (19#3)}
\def\IBIDD#1#2#3{{\it ibid}. {\bf #1}, #2 (20#3)}

\def\APJLL#1#2#3{Astrophys. J. Lett. {\bf #1}, L#2 (20#3)}

\def\JETP#1#2#3{Sov. Phys. JETP {\bf #1}, #2 (19#3)}

\def\JP#1#2#3{J. Phys. A {\bf #1}, #2 (19#3)}

\def\NAT#1#2#3{Nature (London) {\bf #1}, #2 (19#3)}
\def\NATT#1#2#3{Nature (London) {\bf #1}, #2 (20#3)}
\def\NPB#1#2#3{Nucl. Phys. {\bf B#1}, #2 (19#3)}

\def\PLB#1#2#3{Phys. Lett. B {\bf #1}, #2 (19#3)}

\def\PLBold#1#2#3{Phys. Lett. {\bf#1B}, #2 (19#3)}

\def\PRD#1#2#3{Phys. Rev. D {\bf #1}, #2 (19#3)}
\def\PRDD#1#2#3{Phys. Rev. D {\bf #1}, #2 (20#3)}

\def\PRL#1#2#3{Phys. Rev. Lett. {\bf#1}, #2 (19#3)}

\def\PRT#1#2#3{Phys. Rep. {\bf#1}, #2 (19#3)}

\def\PTP#1#2#3{Prog. Theor. Phys. {\bf #1}, #2 (19#3)}

\begin{document}

\twocolumn[\hsize\textwidth\columnwidth\hsize\csname
@twocolumnfalse\endcsname
%%
% OU-TAP 179
%%
\tighten
\draft
\title{Lagrangian evolution of global strings}
\author{Masahide Yamaguchi}
\address{Research Center for the Early Universe, University of Tokyo,
  Tokyo 113-0033, Japan}
\author{Jun'ichi Yokoyama}
\address{Department of Earth and Space Science, Graduate School of
Science, Osaka University, Toyonaka 560-0043, Japan}
%%%

\date{\today}

\maketitle

\begin{abstract}
    We establish a method to trace the Lagrangian evolution of
    extended objects consisting of a multicomponent scalar field in
    terms of a numerical calculation of field equations in three
    dimensional Eulerian meshes. We apply our method to the
    cosmological evolution of global strings and evaluate the energy
    density, peculiar velocity, Lorentz factor, formation rate of
    loops, and emission rate of Nambu-Goldstone (NG) bosons.  We
    confirm the scaling behavior with a number of long strings per
    horizon volume smaller than the case of local strings by a factor
    of $\sim$ 10. The strategy and the method established here are
    applicable to a variety of fields in physics.
\end{abstract}

\pacs{PACS number(s): 98.80.Cq \hspace{5.5cm} RESCEU-11/02,OU-TAP-179}

]

Extended objects consisting of scalar fields such as topological
defects or solitons play important roles in many fields of physical
science ranging from condensed matter physics to cosmology.  Much work
has been done on the cosmological formation of topological defects by
thermal \cite{Kibble} or nonthermal \cite{KVY} phase transitions,
their subsequent evolution \cite{VS}, and cosmological implications of
Q-balls \cite{Coleman}. On the other hand, it has been pointed out
that cosmological defect formation may be realizable in a laboratory
\cite{Zurek}. The phase transition of liquid helium can also be
described in terms of Ginzburg-Landau theory as in the case of
cosmological phase transitions. The order parameter, which is an
expectation value of a scalar field in cosmological phase transition,
is just replaced by a Bose condensate wave function.  Following
Zurek's suggestion, the formation of topological defects has been
studied in $^{3}$He \cite{He3} and $^{4}$He \cite{He4}. In fact,
vortex formation is observed in $^{3}$He. Thus, it is very important
to investigate the formation and subsequent evolution of extended
objects, such as topological defects, produced both in cosmology and
laboratories.

Among various types of topological defects predicted in high energy
physics, strings hold a unique position, which, unlike magnetic
monopoles and domain walls, do not overclose the universe because they
settle down to the scaling solution \cite{Kibble,Kibble2}.  The key
mechanism to achieve it is intercommutation of infinite strings to
lose their energy by producing closed loops, which decay through
radiating gravitational waves for local strings\footnote{The
possibility has also been pointed out that infinite strings lose their
energy by directly emitting massive particles for local strings
\cite{VHS}.} or Nambu-Goldstone (NG) bosons for global strings.

While many grand unified theories may contain local strings, there are
no particle physics motivations for their existence.  Nonetheless they
were extensively studied because it was expected that they might be
the origin of density fluctuations which seeded cosmic large-scale
structures.  In particular, time evolution of local cosmic strings
have been investigated by a number of authors \cite{AT,AT2,BB,AS} who
used the Nambu-Goto action \cite{NG} that applies for infinitely thin
strings. These numerical analyses confirmed the scaling behavior and
the scaling parameter, $\xi$, which is defined as $\xi = \rho_{s}
t^{2} / \mu$ with $\rho_{s}$ energy density of strings and $\mu$
string tension per unit length, converged to $\xi\simeq 10$ in the
radiation dominated universe \cite{AT2,BB,AS}. Recent observations of
anisotropies of cosmic microwave background radiation, however,
disfavor cosmic-string scenario of structure formation \cite{cmb}, and
motivations to study local strings have somewhat diminished.

Global strings, on the contrary, are well motivated in particle physics,
whose existence is predicted by the Peccei-Quinn solution to the strong
CP problem \cite{PQ}. They are produced when the Peccei-Quinn U(1)
symmetry is broken \cite{PQ,VE}. They radiate axions as associated NG
bosons \cite{Davis} which are one of the best candidates of cold dark
matter. Furthermore, they are also realized in condensed matter
physics. Vortices of $^3$He and $^4$He studied in a laboratory are
analogous to global strings \cite{GP}.

In contrast to local strings, energy density of global strings is
dominated by the gradient energy of the NG scalar field rather than
the potential energy of the core, because there are no gauge fields to
cancel the former. As a result they have a long-range interaction and
their dynamics cannot be analyzed with the Nambu-Goto action. It
should be described with the Kalb-Ramond action instead \cite{KR},
which incorporates NG bosons and their coupling with the string core
in addition to the Nambu-Goto action. In this action, however, the
line density of the string is not well defined which has a logarithmic
dependence on the radius.  It is also very difficult to analyze it
numerically. Hence the dynamics of strings governed by the Kalb-Ramond
action has not been clarified in cosmological context yet even though
evolution of axionic strings has very important implications in
cosmology and particle physics. Instead, the results of numerical
simulations of local strings have been used to discuss cosmology of
global strings as well, which may not be justified because evolution
of global defects can be significantly different from that of local
counterparts as seen in the case of local \cite{Preskill} and global
\cite{monopole,Yamaguchi} monopoles.

In order to surmount such a situation, we have investigated evolution of
global strings by solving equation of motion of the complex scalar field
numerically in a three dimensional lattice in the expanding universe
instead of dealing with the Kalb-Ramond action in collaboration with
Kawasaki \cite{YKY,YYK}. In these fully Eulerian simulations, it is
difficult to identify strings and follow their dynamics.  In the
previous publications, whether a lattice contains a string core was
judged by comparing potential energy density at each vertex with a
static cylindrically symmetric solution of a global string. Counting
number of lattices thus identified as containing a string, we estimated
the scaling parameter $\xi$ and found that $\xi \simeq 1$, which is
significantly smaller than the case of local strings. This result has
been criticized by Martins and Shellard \cite{MS,MSM} who claim that we
have obtained incorrectly smaller value of the scaling parameter because
the energy loss of long strings due to direct emission of NG bosons is
much larger in simulation with small dynamic range than in the actual
cosmological context, and that both types of long strings should behave
quantitatively in the same way on cosmological scales.

In order to characterize cosmological evolution of global strings
completely, we need to establish the method to follow cosmological
evolution of global strings from a Lagrangian point of view. The energy
density of long strings, $\rho_{\infty}$ satisfies the following
equation:
\beq
  \frac{d\rho_{\infty}}{dt} = - 2 H (1 + \la v^{2} \ra) \rho_{\infty}
      - \Gamma_{\rm loop}\rho_{\infty}
      - \Gamma_{\rm NG}\rho_{\infty},
  \label{eq:energyloss}
\eeq
where the second and the third terms on the right-hand-side represent
energy loss due to loop formation and direct emission of NG bosons or
axions, respectively, and $\la v^{2} \ra$ denotes average square
velocity of string segments. In the scaling regime, the string network
is characterized by a scale $L$, which is defined as $L \equiv
\sqrt{\mu / \rho_{\infty}}$ and grows with the horizon scale $L
\propto t$. Here $\mu$ is the effective line energy density of a
global string. We can introduce a loop production coefficient $c$ and
an emission coefficient $\kappa$ by
\bea
\Gamma_{\rm loop}\rho_{\infty}= c \frac{\rho_{\infty}}{L},~~~~~
  \Gamma_{\rm NG}\rho_{\infty} 
     = \kappa \frac{\rho_{\infty}}{L}. 
  \label{eq:ckappa}
\eea
Then, the scaling parameter $\xi$ is characterized by these
coefficients as
\beq
  \xi = \lmk \frac{1 - \la v^{2} \ra}{c + \kappa} \rmk^{2}.
  \label{eq:relation}
\eeq
Indeed if $\kappa$ incorrectly turned out to be much larger than $c$,
we would find a smaller value of $\xi$ than it should really be.
Hence it is essential to evaluate contribution of each term in
(\ref{eq:energyloss}) with the help of numerical analysis. For this
purpose we must know how each string segment moves and intercommutes
with each other to trace formation of loops, which is impossible in
conventional Eulerian simulations. Notice that this problem is
ubiquitous because in many fields of physics we must solve equation of
motion in Eulerian meshes, for example, in order to trace time
evolution of solitons like Q-balls \cite{Coleman}.

In the present Rapid Communication, we establish a method to follow
Lagrangian evolution of extended objects in the context of Eulerian
calculation of scalar field equations on the lattice. As a specific
example, we concentrate on global strings here, but the strategy and
the method established here directly apply to other situations in
physics with some proper modifications.

We express the complex scalar field, which constitute global strings, in
terms of two real scalar fields $\phi_{a}(x)$ ($a = 1, 2$) and define
their Lagrangian density as
\beq
  \CL[\phi_{a}] = \frac12 g_{\mu\nu}
                   (\del^{\mu}\phi_{a})(\del^{\nu}\phi_{a})
                    - V[\phi_{a},T],
  \label{eq:lagrangian}
\eeq 
where $g_{\mu\nu}$ is the flat Robertson-Walker metric. The
finite-temperature effective potential $V[\phi_{a},T]$ is taken as
\bea
  V[\phi_{a},T] &=& \frac{\lambda}{4}(\phi^{2} - \sigma^2)^2 
                 + \frac{\lambda}{6}T^2\phi^{2},
~~~\phi^2\equiv\phi_{1}^{2} + \phi_{2}^{2}  
  \label{eq:potential}
\eea
which exhibits a second-order phase transition with the critical
temperature $T_{c} =\sqrt{3}\sigma$ which produces global strings by
breaking a global U(1) symmetry.

Equations of motion for the scalar fields are given by
\beq
  \ddot{\phi_{a}}(x) + 3H\dot{\phi_{a}}(x) 
    - \frac{1}{R(t)^2}\nabla^2\phi_{a}(x)
      = - \frac{\del V}{\del \phi_{a}},
  \label{eq:EOM}
\eeq
where a dot denotes time derivative and $R(t)$ is the cosmic scale
factor. In the radiation dominated universe, the Hubble parameter $H =
\dot R(t)/R(t)$ and cosmic time $t$ are given by
\bea
  H^2 = \frac{8\pi}{3 M_{\rm pl}^2} \frac{\pi^2}{30} g_{*} T^4,
   ~~~~~
  t = \frac{1}{2H},
  \label{eq:hubble}
\eea
respectively, where $M_{\rm pl} = 1.2 \times 10^{19}$GeV is the Planck
mass and $g_{*}$ is the total number of degrees of freedom for the
relativistic particles. For the ease of numerical calculations we take
$\sigma =0.1(45/16\pi^3g_*)^{1/2}M_{\rm pl}$ and $\lambda = 0.08$ but
the results are insensitive to these choices. We start numerical
simulation at the temperature $T_{i} = 2 T_{c}$ corresponding to
$t_{i} = t_{c}/4$ and adopt as an initial condition the thermal
equilibrium state with a mass squared equal to the inverse curvature
of the potential at that time. We have simulated the system from ten
different thermal initial conditions under the periodic boundary
condition. The number of lattices is $128^{3}$ with the lattice
spacing $\delta x = \sqrt{3 t t_{i}} / 8$ and the time step $\delta t
= 0.01t_i$. Thus, the box size is nearly equal to horizon volume
$(H^{-1})^{3}$ and the lattice spacing to a core size of a string
$\delta = 1/(\sqrt{2\lambda}\sigma)$ at the final time $t_{f} = 200
t_{i}$.

The above setup of calculation is the same as our previous fully
Eulerian simulations \cite{YKY,YYK}. In order to reproduce Lagrangian
evolution of string segments in the Eulerian meshes, we develop a
different method to identify string cores from scalar field
configurations, because the previous method \cite{YKY,YYK} was
inadequate for configurations with large curvature such as small loops
and it was impossible to find more correct position of string core in
a box beyond the lattice spacing. Here we use a new two-step algorithm
to identify the locus of a string more accurately. The first step is
to find plaquettes which a string penetrates. This is done in terms of
the Vachaspati-Vilenkin algorithm \cite{VV} by monitoring phase
rotation around each square. Then using the value of $\phi_a$ at each
vertex of a plaquette penetrated by a string, we linearly interpolate
$\phi_a(x)$ to calculate the position where both $\phi_1(x)$ and
$\phi_2(x)$ vanish. We can thus find where a string penetrate in each
plaquette and find more accurate trajectory of a string by connecting
these points.

The next and the more important task to obtain Lagrangian evolution of
string network, that has been missing in the previous Eulerian analyses,
is to find velocity of each string segment. Since motion tangential to a
string is a gauge mode, we should evaluate velocity normal to it.
Suppose that a string core exists at a point $\vect x_0$ at time $t_0$,
namely, $\phi_1(\vex_0,t_0)=\phi_2(\vex_0,t_0)=0$. In order to estimate
where this string segment moves $\Delta t$ later, we expand scalar
fields $\phi_{a}(\vect x,t_0+\Delta t)$ around $\phi_{a}(\vect
x_{0},t_{0})$ up to the first order,
\bea
  \phi_{a}(\vect x,t_0+\Delta t) &\cong& \phi_{a}(\vect x_{0},t_{0}) 
      + \nabla\phi_{a}(\vect x_{0},t_{0}) \cdot (\vect x-\vect x_{0}) 
       \non \\
      && + \dot\phi_{a}(\vect x_{0},t_{0})\Delta t
      \qquad ( a = 1, 2).
  \label{eq:expansion}
\eea
Again the loci of string core at $t=t_0+\Delta t$ are given by the
line where both scalar fields $\phi_{a}(\vect x,t_0+\Delta t)$ vanish.
That is, under the first-order approximation (\ref{eq:expansion}) it
lies on the intersection of two planes given by
\beq
 \nabla\phi_{a}(\vect x_{0},t_{0}) \cdot (\vect x-\vect x_{0})
 +\dot\phi_{a}(\vect x_{0},t_{0})\Delta t=0,  \label{plane}
\eeq
with $a=1,2$. Now suppose that the line normal to the string segment
at $(\vect x,t_0)$ intersect with the trajectory of string at
$t=t_0+\Delta t$ at $\vect x =\vect x_{l}(\vect x_0,t_0,\Delta t)$.
Then, making use of the fact that $\vect x_{l}$ satisfies Eq.\ 
(\ref{plane}), we can easily calculate the velocity of this string
segment, and its magnitude is given by
\bea
  \vect{v}(\vect x_0,t_0) &=& \lim_{\Delta t \longrightarrow 0}
\frac{\vect x_{l}(\vect x_0,t_0,\Delta t) - \vect x_{0}}{\Delta t}
\nonumber \\
    &=& \frac{\dot\phi_{1}\nabla\phi_{2} - \dot\phi_{2}\nabla\phi_{1}}
           {|\nabla\phi_{1} \times \nabla\phi_{2}|}.
  \label{eq:velocity}
\eea    
One should note that the formula obtained here applies to all
soliton-like objects consisting of a field configuration with some
proper modifications.

Now that we have developed a method to identify the location and
velocity of strings at each time, the Lagrangian evolution of the string
network can be traced just as in the case of numerical simulation of the
evolution of local strings based on the Nambu-Goto action. In fact, our
simulation, being based on scalar field equations, contains even more
information, that is, we can find fate of intersecting string segments,
namely, whether they reconnect or simply pass through each other,
without assigning the probability of reconnection by hand unlike in the
case of simulations based on the Nambu-Goto action. We can therefore
calculate formation rate and spectrum of string loops without
ambiguities.

Thus we can calculate all the terms of eq.\ (\ref{eq:energyloss})
directly from simulation data except for the last term, which can be
evaluated by use of this equation itself. As pointed out in
\cite{MSM}, we should use $\mu=\gamma\mu_s$ as the energy per length
to calculate the energy density of long strings, $\rho_{\infty}$. Here
$\gamma$ is the average Lorentz factor and $\mu_s$ is the line density
of a static string given by $\mu_s\simeq
2\pi\sigma^2\ln(t/(\delta\xi^{1/2}))$.

\begin{figure}[htb]
  \begin{center}
    \leavevmode\psfig{figure=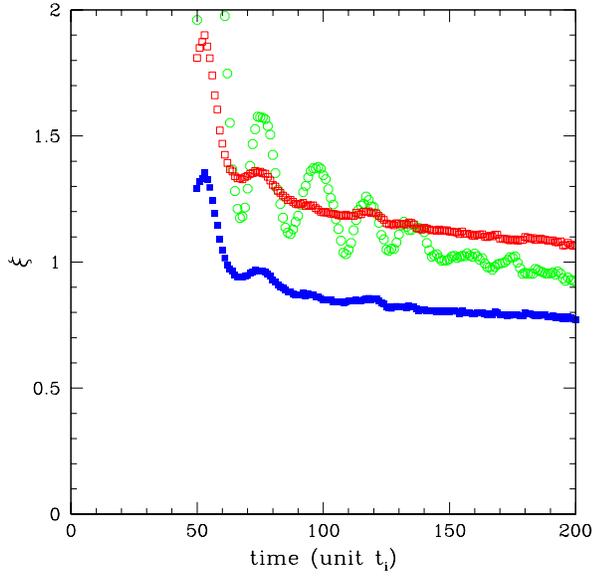,width=8cm}
  \end{center}
  \caption{Time development of $\xi$ is depicted. Blank circles
  represent time development of $\xi$ for the identification method
  done in Refs. {\protect \cite{YKY,YYK}}. Filled squares represent
  the new identification method. Blank squares represent the
  identification method based on the Vachaspati-Vilenkin algorithm.}
  \label{fig:xi}
\end{figure}

Figure \ref{fig:xi} depicts time evolution of $\xi$ with three
different identification methods of strings. Filled squares represent
results of our new identification scheme, while blank circles and
blank squares correspond to our previous method \cite{YKY,YYK} and
Vachaspati-Vilenkin algorithm, which was adopted in \cite{MSM},
respectively.  Both methods estimate total string length by simply
counting the number of boxes penetrated by a string.  As is easily
seen, our previous method slightly overestimates $\xi$ and
Vachaspati-Vilenkin approach overestimates $\xi$ by a factor of $1.4$.
We confirm that $\xi$ becomes constant after some relaxation period
and find that the more correct value of the scaling parameter in the
radiation dominated era is given by $\xi \simeq 0.80$.

%\begin{figure}[htb]
%  \begin{center}
%    \leavevmode\psfig{figure=statistics14.ps,width=8cm}
%  \end{center}
%  \caption{Time development of $\xi$ is depicted. Blank circles
%  represent time development of $\xi$ for identification method done
%  in Ref. {\protect \cite{YKY,YYK}}. Filled squares for new
%  identification method. Blank squares for identification method based
%  on the Vachaspati-Vilenkin algorithm.}
%  \label{fig:xi}
%\end{figure}

\begin{figure}[htb]
  \begin{center}
    \leavevmode\psfig{figure=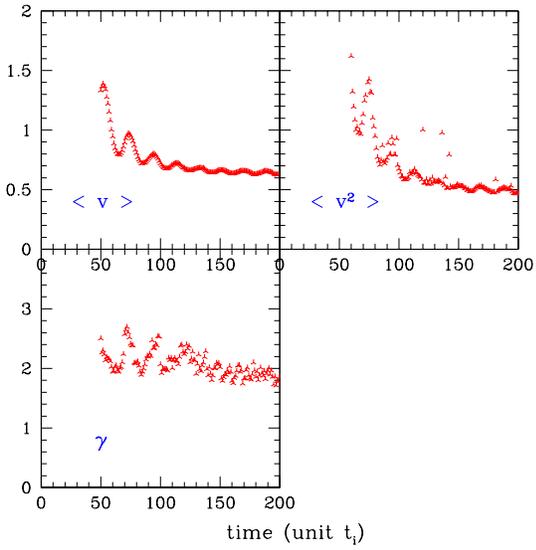,width=8cm}
  \end{center}
  \caption{Time development of average velocity, an average of
  velocity squared, and an average Lorentz factor of global strings is
  shown.}
  \label{fig:tv}
\end{figure}

Figures \ref{fig:tv} depict time evolution of average velocity,
average velocity squared, and average Lorentz factor. The first two
quantities are also found to relax to constant values after some
relaxation time, namely, $\la v \ra \simeq 0.65$ and $\la v^{2} \ra
\simeq 0.50 \gg \la v \ra^{2}$.  The average Lorentz factor, however,
has a large scatter in time, although long-time average is fairly
constant with $\bar{\gamma} \simeq 1.8$.  Since string segments moving
with a speed close to light velocity have extremely large Lorentz
factor and push up the average dramatically, fluctuation in the number
of such string segments results in such large scatter. This also
explains why the average Lorentz factor $\gamma=1.6-2.0$ is larger
than $1/\sqrt{1 - \la v^{2} \ra}$ and $ 1/\sqrt{1 - \la v \ra^{2}}$.
Thus the energy per unit length of a string is enhanced by a factor
$\bar{\gamma} \simeq 1.8$.

Calculating the length of all infinite strings and that of loops at
each time step and comparing with those at the preceding time step,
the loop production parameter is found to be $c = 0.43 - 0.53$. Then
taking account of the relation (\ref{eq:relation}), the emission
parameter $\kappa$ is calculated as $\kappa = 0.03 - 0.13$. We
therefore conclude that the direct emission of NG bosons is a
subdominant channel of energy loss of long strings even in the case
cosmic horizon scale is not much larger than the string width due to
the limitation of the dynamic range of numerical simulations. Thus,
even if $\kappa$ might become significantly smaller in the
cosmological situation because it is proportional to $\ln(t/\delta)$
as pointed out in \cite{MS,MSM}, the scaling parameter $\xi$ would not
increase to coincide with that of local strings. The result of our
simulation with $\ln(t/\delta) \sim 5$ implies that $\kappa \lesssim
0.01$ in the cosmological situations.  Inserting this new value of
$\kappa$ into the relation (\ref{eq:relation}) together with $c$,
$\kappa$, and $\la v^{2} \ra$ yields $\xi = 0.9 - 1.4$ in the
cosmological situations, which is still significantly smaller than the
case of local strings. Hence, it is confirmed that although global
strings relax to the scaling solution just as local strings, their
number density is much smaller than the local ones. This is mainly
because global strings intercommute more often and $\la v^{2} \ra$ is
larger due to long-range attractive forces between strings. Finally,
we apply our results into the constraint of the breaking scale $f_{a}$
of the Peccei-Quinn U(1) symmetry. Then, in accordance with Ref.
\cite{YKY}, $f_{a}$ is constrained as $f_{a} \lesssim (0.20 - 1.6)
\times 10^{12}$ GeV for the normalized Hubble parameter $h = 0.7$ with
$\xi = 0.9 -1.4$ and $\gamma \simeq 1.6 - 2.0$.

In summary, we have developed a new method to trace Lagrangian
evolution of topological defects solving scalar field equation on
three dimensional Eulerian lattices and given all the quantities
characterizing cosmological evolution of the global string network,
namely, energy density, peculiar velocity, Lorentz factor,
intercommutation rate, and emission rate of NG bosons. All these
quantities can be obtained in Eulerian simulations only after
establishing the method given in this Rapid Communication, which
enables us to extract Lagrangian information from simulations done in
Eulerian meshes. Our method is directly applicable to situations in a
variety of fields in physics.

M.Y. and J.Y. are partially supported by the Japanese Grant-in-Aid for
Scientific Research from the Ministry of Education, Culture, Sports,
Science, and Technology, Nos.\ 12-08555(MY) and 13640285(JY).

%\end{document}

\end{document}